\documentclass[journal,onecolumn,12pt]{IEEEtran}

\IEEEoverridecommandlockouts
\IEEEoverridecommandlockouts
\newcommand{\nonl}{\renewcommand{\nl}{\let\nl\oldnl}}% Remove line number for one line
\usepackage[utf8]{inputenc} 
\usepackage{url}
\usepackage{ifthen}
\usepackage{cite}
\usepackage[cmex10]{amsmath} % Use the [cmex10] option to ensure complicance
\usepackage{textcomp}
\usepackage[utf8]{inputenc}   % UTF-8 support
\usepackage[cmex10]{amsmath}
\usepackage{amsmath}          % Math environments
\usepackage{amssymb}          % Extra symbols (\triangleq, \mathbb, etc.)
\usepackage{amsfonts}         % Blackboard bold (\mathbb)
\usepackage{mathtools}        % Small math refinements
\usepackage{float}
\usepackage{bbm}
\let\labelindent\relax
\usepackage{enumitem}
\usepackage{algorithm}
\usepackage{graphicx}
\usepackage{algpseudocode} % Required for \ElsIf
\usepackage{amsthm}
\usepackage{ifthen}
\usepackage[table]{xcolor}
\usepackage{array}
\usepackage{float}
\usepackage{mathtools}
\usepackage{bm}
\usepackage{tikz}
\usetikzlibrary{arrows.meta,positioning,shapes.multipart,shapes.geometric}
\usepackage{etoolbox}
\usepackage{amsmath,amssymb}
\usepackage{xcolor}
\usetikzlibrary{decorations.pathreplacing,calligraphy}

\newcommand{\Fb}{\mathbb{F}} % field
\newcommand{\rank}{\mathsf{rank}} 

\newcommand{\Ber}{\mathsf{Ber}}

\newcommand{\row}[2]{#1[{#2},:]} 

\newcommand{\col}[2]{#1[:,{#2}]}

\newcommand{\w}{\text{wt}_H}

\newcommand{\s}[1]{\left\langle #1 \right\rangle}

\newcommand{\Dens}{\mathsf{SD}}
\newcommand{\MDens}{\mathsf{M}\text{-}\mathsf{DenSD}}

\usepackage[capitalise]{cleveref}

\newtheorem{theorem}{Theorem}
\newtheorem{corollary}{Corollary}

\newtheorem{remark}{Remark}
\newtheorem{definition}{Definition}

\def \sm {\small}

\def \sc {\scriptsize}

\usepackage{tikz,tabularx}
\usetikzlibrary{decorations.pathreplacing}
\tikzstyle{every picture}+=[remember picture]

\usetikzlibrary{shapes,arrows,shadows}
\usepgflibrary{patterns}
\usetikzlibrary{patterns}
\usetikzlibrary{shapes.geometric}
\usetikzlibrary{arrows}

\usepackage{tikz,tabularx}
\usetikzlibrary{decorations.pathreplacing}
\tikzstyle{every picture}+=[remember picture]

\usetikzlibrary{decorations.markings}

\begin{document}
\title{Blind Identification of Channel Codes:\\ A Subspace-Coding Approach} 

% %%% Single author, or several authors with same affiliation:
% \author{%
%  \IEEEauthorblockN{Author 1 and Author 2}
% \IEEEauthorblockA{Department of Statistics and Data Science\\
%                    University 1\\
 %                   City 1\\
  %                  Email: author1@university1.edu}% }

%%% Several authors with up to three affiliations:

\author{Pramod Singh, Prasad Krishnan, Arti Yardi % <-this % stops a space
%\thanks{\hrule}%

\thanks{Pramod Singh, Dr. Prasad Krishnan and Dr. Arti Yardi are with the Signal Processing and Communications Research Center, International Institute of Information Technology, Hyderabad, 500032, India (email: $\{$pramod.singh@research.,  prasad.krishnan@,arti.yardi@$\}$iiit.ac.in). 
% Acknowledgment:  Dr. Krishnan acknowledges support from ANRF-SERB project CRG/2023/008696.
}
}

\maketitle

\allowdisplaybreaks % Command for 

%%=====================================================
\begin{abstract}
% THIS PAPER IS ELIGIBLE FOR THE STUDENT PAPER AWARD. 
% THIS PAPER IS ELIGIBLE FOR THE STUDENT PAPER AWARD \\
The problem of blind identification of channel codes at a receiver involves identifying a code chosen by a transmitter from a known code-family, by observing the transmitted codewords through the channel. Most existing approaches for code-identification are contingent upon the codes in the family having some special structure, and are often computationally expensive otherwise. Further, rigorous analytical guarantees on the performance of these existing techniques are largely absent. This work presents a new method for code-identification on the binary symmetric channel (BSC), inspired by the framework of subspace codes for operator channels, carefully combining principles of hamming-metric and subspace-metric decoding. We refer to this method as the \textit{minimum denoised subspace discrepancy decoder}. We present theoretical  guarantees for code-identification using this decoder, for bounded-weight errors, and also present a bound on the probability of error when used on the BSC. Simulations demonstrate the improved performance of our decoder for random linear codes beyond existing general-purpose techniques, across most channel conditions and even with a limited number of received vectors. 
%%%
\end{abstract}
%TO BE INCLUDED LATER 
% \textit{\small IF NEEDED Due to space restrictions, this submission is a shorter version. The extended version of this work is available on ArXiv \cite{long_version}, and contains the missing proofs, additional results, examples, and figures.}
%%%

%%==================================================
%%==================================================
%\section*{Comments to delete}
%\input{Comments_Temp}

%%==================================================
%%==================================================
\section{Introduction}
\label{sec:intro}

In \textit{blind identification of channel codes}, one considers the setup of coded communication on a channel in which the receiver does not know the exact identity of the channel code being used and the goal is to identify it only using the received data~\cite{Valembois2001}. Such a scenario arises in the context of electronic warfare applications, when intercepting communications from an adversary. Alternatively, in settings involving adaptive modulation and coding~\cite{Goldsmith_Book}, the receiver may be unaware of the code chosen by the transmitter, potentially avoiding additional overhead bits. This setup also has a close connection to cryptanalysis of McEliece cryptosystem~\cite{McEliece_Tillich_2015}.

%\cite{McEliece_Tillich_Polar_2016}

When the receiver has no knowledge of the family of codes from which the transmitter chooses its channel code, the code-identification problem is known to be NP-hard~\cite{Valembois2001}. Thus, typically, researchers assume a specific family of the codes that is used by the transmitter, such as convolutional codes~\cite{Moosavi_journal, Convo_Tillich_2014, Convo_Rank_2023}, low-density parity-check (LDPC) codes~\cite{Cluzeau2006, Xia_LDPC_2013, C1_C2_LDPC_Cosine_2020, C1_C2_LDPC_KL_2024}, and cyclic codes~\cite{Zhou2013_Entropy_new, Arti_TCOMM_2016, Ding2023joint} and design code-identification algorithms that are tailored to a particular code family. Several variations of this problem where the aim is to identify the synchronization of received data, identify the underlying family of the code, or estimate parameters such as length and dimension of the code have also been studied~\cite{CluF2009, SicotHouBaJournal2009, Swaminathan_ISITA_2018, Swaninathan_classi_2017}. 
%explored in the literature~\cite{CluF2009, SicotHouBaJournal2009, Swaminathan_ISITA_2018, Swaninathan_classi_2017}. 
%In this work, we focus on the situation when the transmitter chooses a code from a family of binary linear block codes of the same length (but assume no specific code structure such as cyclic or LDPC) and assume that the noise is introduced by the binary symmetric channel (BSC).

%\tcr{issues with existing methods.. one line.. baseline which?}

While a variety of such algorithms are available in the literature, we would like to highlight that most works 
focus on a particular channel code family and rely on simulations to demonstrate the performance of these algorithms, while the analytical aspects are often minimal. 
Authors of \cite{CluTi2008} provide an information-theoretic approach for this problem where they find a bound on the number of received vectors required for code-identification. However, these bounds are computable only for some specific code families (such as single parity-check codes and regular LDPC codes), while in general they are not computable. 
Authors of \cite{Arti_ISIT_2014, Arti_PhD_Thesis} study optimal likelihood ratio test (LRT) and generalized likelihood ratio test (GLRT) for this problem. However, for these approaches providing analytical performance guarantees is computationally challenging.
%While a variety of such algorithms are available in the literature, we would like to highlight here that most works focus on a particular channel code family.
%
An \textit{inner-product method}~\cite{Valembois2001,Chabot_2007, Xia_LDPC_2013, C1_C2_LDPC_Cosine_2020} is introduced in the literature which can work for any code family.
%and also has low complexity. 
However, this method has a drawback that it works well only when there are \textit{low weight} codewords in the dual code and in general, need a large number of received vectors for code identification.
Thus, a general framework for studying this code-identification problem with the simultaneous requirements that - (a) it applies to general code families (b) has efficient code-identification algorithms, and (c) has analytical guarantees for the performance of such algorithms -  seems to be missing in the literature, to the best of our knowledge. 

A major contribution of this work is to present such a general framework using which the code-identification problem can be analyzed for any code family. Towards this, we establish a novel connection between the \textit{code-identification problem} and the \textit{subspace coding problem} studied in the domain of network coding~\cite{Koetter2008Coding,silva_ksishch_RankmetricApproach,yeung2006network_part1, cai2006network_partII, analog_subspace_coding, Silva_Ksischang_TIT_2010_CommOverMatrixChannels, Capacity_FF_Matrix_2025}. In the subspace coding problem, the goal of the transmitter is to convey a subspace (or a channel code) to a receiver via an \textit{operator channel}\footnote{Examples of an operator channel include the non-coherent multiple-input multiple-output channel where channel matrices are unknown~\cite{Zheng2002}, and a matrix channel that arises in a random network coding~\cite{Koetter2008Coding}.}. 
We observe that the two problems share identical goals, although their underlying channels differ. 
We would like to highlight here that while both  problems have been studied extensively in the literature, yet this natural connection between the two problems remains largely unexplored, to the best of our knowledge.   
In our work, by exploiting this connection, we present a new code-identification method which has both theoretical guarantees and good performance on the binary symmetric channel (BSC). We believe that our subspace-coding based framework will provide a fresh perspective for studying various problems related to the code-identification setup as well.

%establishing a connection between them will let one to study the code-identification problem under a fresh and novel framework that the ideas from subspace coding literature may provide. 
%\tcr{By observing a natural connection between subspace doing problem and code-identification problem .. by exploiting this connection , present a new decoder which has both theoretical guarantees and good performance on the BSC ... decoding of subspace codes.. }

% \pk{This para can be commented out in short version} 
In the subspace-coding literature, the notion of \textit{channel discrepancy} between the channel-input and channel-output has been popularly used. Specifically, for general adversarial channels, the authors of \cite{silva_ksischang_onmetricsTIT} define discrepancy as a function that quantifies the adversary’s effort to transform the transmitted symbol to the received symbol. These ideas have enabled the reinterpretation of various models of network coding, including coherent network coding \cite{yeung2006network_part1,cai2006network_partII}
and non-coherent network coding \cite{Koetter2008Coding}, bringing all of these under a single roof. 
Inspired by such an approach, for our code-identification problem we define a new notion of discrepancy to connect the random errors of the BSC with the subspace distance errors that arise in subspace-coding.

To summarize, in this work, we focus on the situation when the transmitter chooses a code from a family of linear block codes of the same length (but assume no specific code structure such as cyclic or LDPC). The codewords are transmitted via the BSC
%binary symmetric channel (BSC) 
and the receiver's goal is to estimate the transmitter's chosen code using the noise-affected data. 
%For this setup, 
The contributions and organization of this work are: 
%summarized below:
%
\begin{itemize}[leftmargin=*]
\item Section \ref{sec:sysmodelandprelim} reviews the system model for blind identification of channel codes, and provides a brief review of relevant algorithms in the literature. We also present the relationship between the decoding of subspace codes on operator channels via the subspace distance metric, and the blind code-identification problem. 
\item In Section \ref{sec:newdecoder}, we present a new notion of a \textit{denoised subspace discrepancy} between the received noisy data and the each code in the family, which captures the distinct features of the code-identification setting better than the subspace distance %termed as \textit{Subspace based Discrepancy} 
(see \cref{def:disc}).  
Using this new discrepancy measure, we define a new code-identification method, which we refer to as the \textit{minimum denoised subspace discrepancy} ($\MDens$) \textit{decoder}  (see \eqref{eq: mdd}). We prove the guaranteed error correction capability of this decoder, under reasonable constraints (see Theorems~\ref{Theorem_MdenSD_Correct}, \ref{Theorem_MSDD_Correct_2}). Subsection \ref{subsec:improveddecoder} provides a further improved version of the $\MDens$ decoder. We also provide a analytical bound on the probability of error of the improved $\MDens$ decoder in Subsection \ref{subsec:errorprobbound}. 
%
% \item For our proposed code-identification method, we provide the necessary conditions for the correct code-identification and also study the guaranteed error correction capability . \tcb{@Prasad: Can you review+edit this?}
%
\item In Section \ref{sec:Simulations}, we provide comparisons between our proposed method with the baseline literature method (see \cref{subsection_literature_review}). We observe that our method performs better than the literature method for higher values of cross-over probability and also when the number of received vectors is limited (see \cref{fig:comparison}). 
\end{itemize}
%
%\subsubsection{Notation} 
\textit{Notation:} For a positive integer $n$, let $[n] = \{1,2,\ldots,n\}$. The empty set is denoted by $\emptyset$.
For $x \in \mathbb{R}$, define $(x)_{+} \triangleq \max\{0,x\}$. $\Fb_2$ denotes the finite field with $2$ elements, while $\Fb$ denotes a generic finite field.  
Matrices are denoted by capital letters. For any matrix $A$, its rank and its row space are denoted as $\rank(A)$ and $\s{A}$, respectively. Let $\row{A}{R}$ denote the submatrix formed by the  rows of $A$ indexed by set $R$. The $r$th row and column of $A$ are denoted by $\row{A}{r}$ and $\col{A}{r}$, respectively. The sum of two subspaces $W_1$ and $W_2$ is written as $W_1+W_2$. The dimension of a subspace $U$ is denoted by $\dim(U)$. For $p\in[0,1]$, $\Ber(p)$ denotes the Bernoulli distribution with parameter $p$.

%%=====================================================
%%=====================================================
%
\section{System Model and Preliminaries}
\label{sec:sysmodelandprelim}

%%=====================================================
%=================================
%
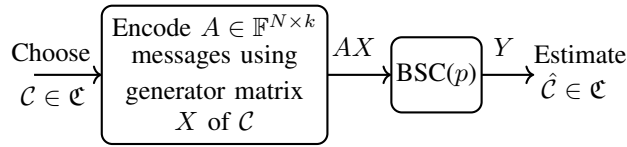
\begin{figure}[htp]

\begin{center}

\begin{tikzpicture}

\node [left] at (0-0.4,0.3) {\sm Choose};
\node [left] at (-0.08-0.4,0.32-0.6) {\sm ${\cal C} \in {\mathfrak C}$};

\draw [->,thick] (-1-0.3,0) -- (0-0.4,0);

%---------------
\draw [thick, rounded corners] (0-0.4,-0.92) rectangle (2.6,0.92);
\node [above] at (1.1,-0.05+0.4) {\sm Encode $A \in \mathbb{F}^{N \times k}$};
\node [above] at (1.1,-0.06) {\sm messages using};
\node [below] at (1.1,0.06) {\sm generator matrix};
\node [below] at (1.1,0.06-0.4) {\sm $X$ of $\cal C$};

%---------------
\draw [->,thick] (2.6,0) -- (2.6+0.85,0);
\node [above] at (2.6+0.35,0.1) {\sm $AX$};

\draw [thick, rounded corners] (2.6+0.85,-0.5) rectangle (2.6+0.85+1.2,0.5);
\node [above] at (2.6+0.85+0.6,-0.3) {\sm BSC($p$)};

%---------------
\draw [->,thick] (2.6+0.85+1.2,0) -- (2.6+0.85+1.2+0.7,0);

\node [above] at (2.6+0.65+1.2+0.5,0.1) {\sm $Y$};

\node [right] at (2+1+1.2+1,0.28) {\sm Estimate};
\node [right] at (2.1+1+1.2+1,-0.15) {\sm $\hat{\cal C} \in {\mathfrak C}$};

\end{tikzpicture}

\end{center}

\caption{Code Identification on BSC$(p)$}

\label{Figure_system_model_intro_identification}

\end{figure}

In this section, we present the system model for the code-identification problem on the binary symmetric channel $BSC(p)$, where $p\in[0,0.5]$ is the cross-over probability of the channel.

Let $\mathfrak{C}=\{{\cal C}_i:i\in[M]\}$ denote a collection of $M$ binary linear codes of length $n$, known to both parties in the communication channel.  The transmitter picks a code ${\cal C}$ out of this collection uniformly at random, and transmits $N$ codewords, each of which is chosen independently and uniformly at random from the codewords in ${\cal C}$. Let $\dim({\cal C}_i)=k_i$, for $i\in[M]$. Let $X_i\in \Fb_2^{k_i\times n}:i\in[M]$ denote generator matrices of the codes ${\cal C}_i:i\in[M]$ respectively. Then, the transmitted $N$ codewords can be written as the matrix $AX$, where $X$ is the generator of ${\cal C}$ and $A\in \Fb_2^{N\times k}$ is the matrix of information vectors chosen uniformly at random. 

The received matrix at the receiver is written as $Y=AX+E$, where each entry of the error matrix $E\in \Fb_2^{N\times n}$ is generated independently according to $\Ber(p)$. In the \textit{code-identification problem}, the decoder's goal is to identify the code chosen for transmission by the sender from the received matrix $Y$. Thus, a decoder is a function $g:\Fb_2^{N\times n}\to \mathfrak{C}$, where we denote the estimate $g(Y)$ as $\hat{{\cal C}}$. We seek to design a decoder which minimizes the probability of identification error, denoted as $P_E(g)\triangleq \Pr(\hat{\cal C}\neq {\cal C}).$

In Section \ref{sec:newdecoder}, we propose a new decoder by combining the Hamming metric and the subspace-distance metric. We now give a brief overview of the subspace distance and its basic properties.

%------------------------
\subsection{Code-Identification: Literature Review}\label{subsetion:lit review} 
\label{subsection_literature_review}
In the literature, most methods
focus on a particular channel code family. In our work, since we consider the situation when the transmitter code can be any arbitrary code, for this literature review we restrict our attention to code-identification methods that can be applied to arbitrary  channel code families.

Authors of \cite{Arti_ISIT_2014} studied the optimal likelihood ratio tests (LRT) for code-identification. However, a major drawback of this method is that the computation of the likelihoods required for the LRT becomes computationally prohibitive as the code dimension increases~\cite{Arti_ISIT_2014}, making its implementation computationally challenging.
In \cite{Arti_PhD_Thesis}, the authors study a method based on the generalized likelihood ratio tests (GLRT). However, this approach requires one to find the maximum likelihood estimates of the received vectors, which would be computationally challenging for random codes. 
An information-theoretic approach to the code-identification problem was presented in \cite{CluTi2008}.
Specifically, the authors prove that 
for code-identification between $M$ possible codes chosen from a general code family, $\Theta(\log_2 M)$ received vectors are required. % (or at least $\log_2 M$ received codewords are required). 

Authors also provide analytical expressions for the bounds on the number of the received codewords  required. However, these bounds are computable only for some specific code families (such as single parity-check codes and regular LDPC codes), while in general they are not computable.
Authors of \cite{Valembois2001, Chabot_2007} have proposed an \textit{inner-product method}. %
This method has been popularly adapted for a wide variety of code families as well~\cite{Cluzeau2006, Moosavi_journal, CluF2009, Xia_LDPC_2013, C1_C2_LDPC_Cosine_2020}. 
% Moosavi_journal, Xia_LDPC_2013
% \tcb{Owing to the computational complexity of the methods discussed as above, for the baseline comparison we focus on the inner-product method, described next.}
%

In this \textit{inner-product method}, one needs to first find a minimum weight vector $\mathbf{h} \in \mathcal{C}_i^{\perp} \setminus \mathcal{C}_j^{\perp}$, for each $\mathcal{C}_i,\mathcal{C}_j \in \mathfrak{C}$ and $\mathcal{C}_i \neq \mathcal{C}_j$.
Then, for the received $Y\in \Fb_2^{N\times n}$, one computes the inner-product $Y\mathbf{h}^T\in \Fb_2^N$. Let $Z$ be the random variable corresponding to the number of one's in $Y\mathbf{h}^T$.
It is known that $Z$ follows the binomial distribution with parameters $\theta_i$ and $\theta_j$ when $Y$ is a noise-affected version of codes $\mathcal{C}_i$ and $\mathcal{C}_j$ respectively, where $\theta_i$ and $\theta_j$ are given by~\cite{Valembois2001, Chabot_2007}
\begin{align}
\theta_i = 0.5 - 0.5(1 - 2p)^{\w{\mathbf{(h)}}} \mbox{~and ~}\theta_j=0.5.
\label{Eqn_inner_product_method}
\end{align}
The problem of distinguishing between $\mathcal{C}_i$ and $\mathcal{C}_i$ can now be formulated and solved as a hypothesis testing problem. 
It should be noted that a major limitation of inner-product method is the requirement for the existence of a low weight $\mathbf{h} \in \mathcal{C}_i^{\perp} \setminus \mathcal{C}_j^{\perp}$, since this directly affects the performance of hypothesis testing (see \cref{Eqn_inner_product_method}). Further, finding such $\mathbf{h}$ might become computationally prohibitive for random codes. Indeed, it is known \cite{complexityofISD} \cite{Algoforminweight} that finding low weight vectors in ${\mathcal{C}_i}^{\perp} \setminus \mathcal{C}_{j}^{\perp}$ needs $\mathcal{O}(2^{\gamma n})$ search operations, for some constant $\gamma$. Owing to the computational complexity of the other methods for arbitrary (random) codes, we consider the inner product method as the baseline in this work.

%\begin{itemize}
    %\item LRT computationally prohibitive.. need to condition over all possible transmitted codewords.. no performance guaratees in term of number of possible errors that we can tolerate.. without clear performance guarantees.. mentioned in intro
    %\item Surprising that connection to subspace coding has not been made yet
    %\item GLRT computationally prohibitive.. need ML estimates of Y
    %\item We compare with the baseline method
%\end{itemize}

\subsection{Relation to Subspace Codes}
\label{subsec:relationtosubspacecodes}
 The subspace distance (see ~\cite{Koetter2008Coding}, for instance) between two subspaces $U$ and $V$ of an $n$-dimensional vector space $\Fb^n$ is defined as $d_s(U,V) \triangleq \dim(U+V) - \dim(U \cap V)$. Note that $d_s(U,V)=\dim(U) + \dim(V) - 2\dim(U \cap V)$. Further, for any two matrices $A \in \Fb^{m_1 \times n}$ and $B \in \Fb^{m_2 \times n}$, it holds \cite{Koetter2008Coding} that 
 
 {\small \begin{align}
 \label{eq:subdist_matrix}
d_s(\s{A},\s{B})
&= 2 \rank\left( \begin{bmatrix}
A\\
B
\end{bmatrix}\right) - \rank(A) -\rank(B). 
\end{align}
}
It is known \cite{Koetter2008Coding} that the subspace distance is a distance metric on the collection of all subspaces of $\Fb^n$. 

The subspace distance was shown to be crucial \cite{Koetter2008Coding,silva_ksishch_RankmetricApproach,silva_ksischang_onmetricsTIT} in the context of communication on a network employing random linear network coding \cite{Ho_et_alTIT2026_RandomLinearNC}, where the input-output matrix relationship is given by $Y=AX+BZ$ (over an arbitrary finite field $\Fb$), where the matrices $A$, $B$ and $Z$ are part of the channel description, and the matrix $X$ represents the transmitter's message. The matrix $A$ is typically modeled in these works as a random matrix (arising out of random variables representing the linear network coding coefficients in the network \cite{Ho_et_alTIT2026_RandomLinearNC}), resulting in a transformed message $AX$ at the client. The matrix $Z$ represents the error packets injected at the edges of the network, and $B$ represents another random matrix modeling the random linear network coding operations applied to the error packets $Z$. 

By this description of the random network coding framework, it can be seen that the information content in $X$ is essentially `carried' through the channel by the \textit{input subspace} $\s{X}$. The effect of the channel matrices $A,B,Z$ can be captured via the subspace distance between $d_s(\s{X},\s{Y})$, where $\s{Y}$ is the \textit{output subspace}. The resulting equivalent channel is then called an \textit{operator channel} \cite{Koetter2008Coding,silva_ksishch_RankmetricApproach}. Following this model, the message $X$ was designed to be chosen from a collection of generator matrices of subspaces in a \textit{subspace code} \cite{Koetter2008Coding}. An $(M,n)$-subspace code $\mathfrak{S}$ \cite{Koetter2008Coding} over $\Fb$ is essentially a collection of $M$ subspaces of $\Fb^n$. The minimum distance $d_{\min}(\mathfrak{S})$ of a subspace code $\mathfrak{S}$ is the minimum of the pairwise distances between the subspaces in $\mathfrak{S}$, i.e., 
$d^{\mathsf{sub}}_{\min}(\mathfrak{S}) \triangleq \min\limits_{\substack{\mathcal{S}_1,\mathcal{S}_2\in \mathfrak{S}: \mathcal{S}_1 \neq \mathcal{S}_2}} d_s(\mathcal{S}_1, \mathcal{S}_2).$ A natural decoding rule for this channel is based on the subspace distance, given as 
\begin{equation}
\label{eq:MSDD}
\hat{\cal S} = \arg\min_{{\cal S} \in \mathfrak{S}} d_s(\s{Y},{\cal S}),
\end{equation}
where ties are resolved arbitrarily. This decoder is referred to as the \emph{minimum subspace distance decoder (MSD) decoder}. This decoder guarantees the following performance. 
%%%%
\begin{theorem} \cite[Theorem 1, rephrased]{silva_ksishch_RankmetricApproach}
\label{thm:subspacedistancedecoderSilva}
   For the channel model $Y = AX + BZ$, let $X$ be chosen at random from the collection of generator matrices of a subspace code $\mathfrak{S}$ with minimum subspace distance $d^{\mathsf{sub}}$ and $Y$. For some $\rho\geq 0$, suppose $\rank(A) \ge \max_{{\cal S}\in\mathfrak{S}}\dim({\cal S}) - \rho $ and $\rank(BZ) < \frac{d^{\mathsf{sub}}-2\rho}{4},$ then the MSD decoder correctly identifies the transmitted subspace $\s{X}\in \mathfrak{S}$.
\end{theorem}

The similarities between the operator channel framework and the blind code-identification setting are clear. Seen over $\Fb_2$, the matrix-pair $(A,X)$ in the blind code-identification has an identical role as the pair in subspace coding. Note that in the code-identification framework, the additive error matrix $E$ depends on the $BSC(p)$, whereas in the literature on network-coding, the error matrix $BZ$ arises out of edge-error packets modulated by the random network code. Despite this difference, we can potentially employ minimum subspace distance decoding, i.e., we calculate the estimate as $\hat{\cal C} = \arg\min_{{\cal C} \in \mathfrak{C}} d_s(\s{Y},{\cal C})$. Theorem \ref{thm:subspacedistancedecoderSilva} continues to guarantee a correct decoding, however under the condition that $A$ has sufficiently large column-rank and $E$ has sufficiently low rank. However, even for small $p$, the rank of the matrix $E$ obtained from the $BSC(p)$ could be high, violating the constraint on $\rank(E)$ in Theorem \ref{thm:subspacedistancedecoderSilva}. Thus, applying the MSD decoder directly for the code-identification problem gives unsatisfactory results (see Figure \ref{fig:comparison}). Nevertheless, the strong connection between the frameworks raises the question: \textit{Is there a subspace-distance inspired decoder that has good performance for the blind channel code-identification problem?} In Section \ref{sec:newdecoder}, we answer this in the affirmative, presenting a new decoder by employing a simple but carefully thought-out modification of the MSD decoder.

%%=====================================================
%%=====================================================
%
\section{A New Decoder for code-identification}
\label{sec:newdecoder}
In this section, we define a new decoder for the code-identification problem, by carefully combining bounded hamming-distance decoding and the subspace distance metric. We call this the \textit{minimum denoised subspace discrepancy} decoder. We show theoretical code-identification guarantees of this decoder for errors on the BSC with bounded weights. We also provide an analytical bound on the probability of error of this decoder, using these theoretical guarantees. 

Towards these, we need a few definitions. For a given code family $\mathfrak{C}$, we define the quantity $\delta$ as the minimum hamming distance between any pair of distinct codewords between any two codes in the family. Formally, 
\begin{equation}
\label{eqn:definitiondelta}
\delta = \min_{\substack{i, j \in [M] \\ i \neq j}}
  \min\limits_{\substack{\boldsymbol{c}\in \mathcal{C}_i,  \boldsymbol{c'}\in \mathcal{C}_j \\ \boldsymbol{c}\neq \boldsymbol{c'}}} d_H(\boldsymbol{c}, \boldsymbol{c'}).
\end{equation}
\begin{remark}
\label{remark:delta}
    Note that $\delta=\min_{i,j\in [M]}d_{\min}({\cal C}_i+{\cal C}_j)$. This follows from the definition. Further, this also implies that $\delta\leq \min_{i}d_{\min}({\cal C}_i)$. 
\hfill $\square$    
\end{remark}

% \begin{definition}(Channel Discrepancy Function)
% Let $\mathcal{X}$ and $\mathcal{Y}$ denote the channel input and output alphabets, respectively. 
% For a given channel model, we define a \emph{discrepancy function} as a mapping
% $
% \Delta : \mathcal{X} \times \mathcal{Y} \longrightarrow \mathbb{R}_{\ge 0},
% $
% which quantifies the dissimilarity between the transmitted symbol $X \in \mathcal{X}$ and the received symbol $Y \in \mathcal{Y}$. 
% \end{definition}
% Intuitively, $\Delta(X, Y)$ measures how far the channel output $Y$ is from the input $X$.

% \begin{definition}(Inter-Code Hamming Distance $\delta$)\label{def:delta}
% Let $\mathcal{C}_i$ and $\mathcal{C}_j$ be two distinct codewords from the subspace code $\mathfrak{C}$. The minimum Hamming distance between them is defined as:
% \begin{equation}
%     d_H({\cal C}_i, {\cal C}_j) = \min \{ \text{wt}_H(u-v) \mid u \in {\cal C}_i, v \in {\cal C}_j, u \neq v \}.
% \end{equation}
% The overall inter-code Hamming distance of the code $\mathfrak{C}$ is the minimum of these values over all distinct pairs of code:
% \begin{equation}
% \label{eqn:definitiondelta}
% \delta = \min_{\substack{i, j \in [M] \\ i \neq j}}
%   d_H(\mathcal{C}_i, \mathcal{C}_j) \\
% \end{equation}
% \end{definition}

The work \cite{silva_ksischang_onmetricsTIT} defined a general notion of a \textit{channel discrepancy} $\Delta(Y,X)$, to quantify the effort needed by an adversarial channel to produce a particular output $Y$, given an input $X$. Taking inspiration from this, we define a discrepancy function for the blind code identification channel model.
%%% 
\begin{definition}(Denoised Subspace Discrepancy)\label{def:disc}
Given a received matrix $Y \in \mathbb{F}_q^{N \times n}$ and a candidate code $\mathcal{C}_i \in \mathfrak{C}$, we define the denoised subspace discrepancy between $Y$ and ${\cal C}_i$ as
\begin{equation}\label{eqn:SD}
   \Delta_{\Dens}(Y,\mathcal{C}_i) \triangleq d_s(\s{D_i(Y)}, \mathcal{C}_i), 
\end{equation}
where $D_i(Y)$ is an $N\times n$ matrix, which is defined row-wise, for all $r\in[N]$, as follows.

{\small \begin{align}
D_i(Y)[r,:]
= \begin{cases}
    \mathbf{c}, & \text{if}~\exists~\mathbf{c}\in{\cal C}_i~\text{with}~d_H(\mathbf{c}, Y[r,:])< \frac{\delta}{2}\\
    Y[r,:], & \text{otherwise}. 
\end{cases} 
\label{eq:decodedrows}
\end{align}
}
\end{definition}
In other words,  $D_i(Y)$ denotes the matrix obtained by decoding only those rows of $Y$ which are $\lfloor(\delta-1)/2\rfloor$-close to a codeword in ${\cal C}_i$. Essentially, row-$r$ of $D_i(Y)$ can be obtained by attempting to perform bounded distance decoding of row-$r$ of $Y$, with respect to the  the code ${\cal C}_i$, upto $\lfloor(\delta-1)/2\rfloor$ errors. Following  \eqref{eqn:definitiondelta} and  \cref{remark:delta}, we see that in case this decoding is successful, the codeword $\mathbf{c}\in{\cal C}_i$ that is obtained will be unique. If the bounded distance decoding fails, then we retain $Y[r,:]$ as is. Thus, $D_i(Y)$ represents a de-noising operation. Potentially, this operation corrects low-weight errors when ${\cal C}_i$ is the true code. 
 
We now define a new decoder for the blind code-identification problem using the $\Delta_{\mathsf{SD}}$ discrepancy. 
\begin{equation}
    \hat{\mathcal{C}}_{SD} = \arg\min_{\mathcal{C} \in \mathfrak{C}} \Delta_{\Dens}(Y, {\cal C}),
    \label{eq: mdd}
\end{equation}
where we assume that the ties are broken arbitrarily. We will call this decoder \textit{Minimum Denoised Subspace Discrepancy Decoder ($\MDens$ Decoder)}. The intuition behind this decoder is that, once the low-weight errors are corrected, the resultant `denoised' matrix $D_i(Y)$ becomes subspace-close to the code ${\cal C}_i$, if that is indeed the true transmitted code. On the other hand, for a wrong code ${\cal C}_j$ ($j\neq i$), the denoised matrix $D_j(Y)$ remains at a larger subspace distance from ${\cal C}_j$. 
%%%%%
\subsection{Guaranteed Identification Capabilities of the $\MDens$ decoder}
\label{subsec:guaranteesofMdensDecoder}
We now present two main theorems of this work, Theorem \ref{Theorem_MdenSD_Correct} and Theorem \ref{Theorem_MSDD_Correct_2}. These results provide guarantees on the correctness of code-identification by the $\MDens$ decoder employed on the BSC, under conditions including that the error matrix $E$ has low-weight rows, and that sufficiently many `uniquely identifiable' codewords are transmitted by the sender. 

\begin{theorem}
\label{Theorem_MdenSD_Correct}
    Let $\mathfrak{C}=\{{\cal C}_j:j\in[M]\}$ be a collection of codes. Let ${\cal C}_i\in{\mathfrak C}$ be the true code that is transmitted. For some $0\leq \rho\leq N$, suppose the following conditions are true.
    \begin{enumerate}[leftmargin=*]
        \item \underline{Sufficient Rank Condition:} $\rank(AX_i)\ge N -\rho $. 
        \item \underline{Low-weight Errors:} The rows of the error matrix $E$ have weight at most $\lfloor{\frac{(\delta-1)}{2}\rfloor}$. 
        \item \underline{Presence of `uniquely identifiable' codewords:}  For each $j\in [M]\setminus i$, the matrix $AX_i$ has at least $(\dim(\mathcal{C}_i) - \dim(\mathcal{C}_j) + \rho +1)_{+}$ linearly independent rows which are codewords in ${\cal C}_i$ but not in ${\cal C}_j$. 
    \end{enumerate}
     Then, the $\MDens$ decoder is correct, i.e., it returns the right estimate $\hat{\cal C}_{SD}={\cal C}_i$.
\end{theorem}
\begin{IEEEproof}
    For the true transmitted code being $\mathcal{C}_i$, and a given received matrix $Y$, consider the denoised matrix $D_i(Y)$. Thus, $Y=AX_i+E$, where $E$ is the error matrix with row-weights at most $\lfloor \frac{\delta -1}{2}\rfloor$. It follows from \eqref{eq:decodedrows} that $D_i(Y) = AX_i$. Now consider the discrepancy between $Y$ and the true code $\mathcal{C}_i$.
    \begin{align*}
        \Delta_{\Dens}&(Y,\mathcal{C}_i) \\ &= d_s(\s{D_i(Y)}, \mathcal{C}_i ) \\
        % &= \dim(\mathcal{C}_i) + \dim(\s{D_i(Y)}) \\
        % &~~~~- 2 \dim(\s{D_i(Y)} \cap \mathcal{C}_i) \\
        &\overset{}{=} \dim(\mathcal{C}_i) + \dim(\s{AX_i}) - 2\dim(\s{AX_i} \cap \mathcal{C}_i) \\
        &\overset{(a)}{=} \dim(\mathcal{C}_i) - \dim(\s{AX_i}) \\
        &\le (\dim(\mathcal{C}_i) - N + \rho)_+,
    \end{align*}
where $(a)$ follows as $\s{AX_i} \subseteq \mathcal{C}_i$, and the last inequality follows from given condition \textit{1)}.

Now consider the discrepancy between matrix $Y$ and a `wrong' code ${\cal C}_j$ for some $j\neq i$. Suppose that, for some row $r\in [N]$, the $r^{\text{th}}$-row of $AX_i$ lies in ${\cal C}_i$ but not in ${\cal C}_j$. Then, we claim that $D_j(Y)[r,:]=Y[r,:]\notin {\cal C}_j$. Indeed, if this were not the case, then by definition of $D_j(Y)$, we must have $d_H(D_j(Y)[r,:],Y[r,:])<\delta/2$. Since $E[r,:]$ is given (condition \textit{2)}) to be of weight strictly smaller than $\delta/2$, by the triangle inequality, this would mean that $d_H(D_j(Y)[r,:],(AX_i)[r,:]))<\delta$. But the only way this can happen, by definition of $\delta$, is if $D_j(Y)[r,:]=(AX_i)[r,:]\in{\cal C}_i$. This is clearly a contradiction. Hence, $D_j(Y)[r,:]=Y[r,:]\notin {\cal C}_j$. Now, we have
\begin{align*}
    \Delta_{\Dens}(Y,\mathcal{C}_j) &= d_s(\s{D_j(Y)} ,\mathcal{C}_j) \\
    &= \dim(\mathcal{C}_j) -  \dim(\s{D_j(Y)} \cap \mathcal{C}_j) \\
    &\quad 
    + \dim(\s{D_j(Y)}) -  \dim(\s{D_j(Y)} \cap \mathcal{C}_j) \\
    &\overset{(b)}{\ge} \dim(\mathcal{C}_j) - \dim(\s{D_i(Y)} \cap \mathcal{C}_j)\\
    &\quad{}+ (\dim(\mathcal{C}_i) -\dim(\mathcal{C}_j) + \rho +1)_{+}
\end{align*}
Here, $(b)$ follows from the fact that there exist at least $(\dim(\mathcal{C}_i) - \dim(\mathcal{C}_j) +\rho + 1)_{+}$ rows in $AX_i$ which are linearly independent codewords in $\mathcal{C}_i \setminus \mathcal{C}_j$ (condition \textit{3)}). Since, $\dim(\s{D_i(Y)} \cap \mathcal{C}_j) \le \dim(\s{D_i(Y)}) \le N$, we can further lower-bound the right hand side (R.H.S.) of $(b)$ to obtain
\begin{align*}
    &\geq \dim(\mathcal{C}_j) - N + (\dim(\mathcal{C}_i) -\dim(\mathcal{C}_j) + \rho +1)_{+}\\
    &\geq  \dim(\mathcal{C}_j) - N + (\dim(\mathcal{C}_i) -\dim(\mathcal{C}_j) + \rho +1)\\
    & >\Delta_{\Dens}(Y,\mathcal{C}_i). 
\end{align*}
Thus, we have shown that $\Delta_{\Dens}(Y,\mathcal{C}_j)>\Delta_{\Dens}(Y, \mathcal{C}_i) ~\forall~\mathcal{C}_j \in \mathfrak{C}, ~\mathcal{C}_i \ne \mathcal{C}_j$. Hence, the $\MDens$ decoder decodes correctly i.e. it gives the true code $\mathcal{C}_i$ as an output.
\end{IEEEproof}
Observe that Condition \textit{1)} of Theorem \ref{Theorem_MdenSD_Correct} implies that the random matrix $A$ (of size $N\times \dim({\cal C}_i)$) has a row-rank deficiency of at most $\rho$. In other words, Theorem \ref{Theorem_MdenSD_Correct} is essentially capturing the scenario when the number of transmitted codewords $N$ is at most $\dim({\cal C}_i)+\rho$, guaranteeing the $\MDens$ decoder's correct identification in this case, when sufficient `uniquely-identifiable' codewords and low-weight errors are ensured. While Theorem \ref{Theorem_MdenSD_Correct} holds for any code family, the subsequent corollary (which follows directly from Theorem \ref{Theorem_MdenSD_Correct}) highlights the scenario when all codes have the same dimension, i.e., when $\mathfrak{C}$ is an \textit{equi-dimensional code family}.
%%%
\begin{corollary}
\label{corr:rho+1UniqueCodewordsenough}
       Let $\mathfrak{C}=\{{\cal C}_j:j\in[M]\}$ be a collection of codes, where $\dim({\cal C}_j)=k, \forall j\in[M]$ and $\mathcal{C}_i$ be the true transmitted code. If $A$ has rank at least $N-\rho$ (for some $0\leq \rho\leq N$) and the transmitted matrix $AX_i$ has at least $(\rho +1)$ linearly independent rows which are codewords in ${\cal C}_i$ but not in ${\cal C}_{j}$, for each $j\neq i$, then the $\MDens$ decoder is correct, i.e., it returns ${\cal C}_i$ as its output. 
\end{corollary}
%%%
\begin{remark}Corollary \ref{corr:rho+1UniqueCodewordsenough}  indicates that a single `uniquely identifiable' codeword is enough for correct identification, if $A$ has full row-rank ($=N$) and the errors are low weight. It is interesting to note that, this fact, which is rather intuitive, agrees with the observation in \cite{CluTi2008} that a small number of codewords is sufficient for code-identification, provided the channel noise is low.
\hfill $\square$    
\end{remark}

Now, if sufficiently many transmissions are made that enough rank is `accumulated' in the matrix $A$, we must expect that the condition of `uniquely identifiable' codewords is automatically satisfied, since the codes in $\mathfrak{C}$ are distinct and contain such codewords. Formalizing this observation, we obtain Theorem \ref{Theorem_MSDD_Correct_2}, which provides a further guaranteed correct identification via the $\MDens$ decoder, when $N$ is large enough that $\rank(A)$ equals the dimension of of the transmitted code. Standard probabilistic arguments show that this happens exponentially fast as $N$ grows beyond the dimension of the code (see \cite{Ferriera_2013_USEFUL_rankofrandombinarymatrix,Rankresult}, for instance). 
%%%%
\begin{theorem}
\label{Theorem_MSDD_Correct_2}
    Let $\mathfrak{C}=\{{\cal C}_i:i\in[M]\}$ be a collection of codes, where $\dim({\cal C}_i)=k_i, \forall i\in[M]$, such that no code in $\mathfrak{C}$ is contained completely within another. Let ${\cal C}_i\in{\mathfrak C}$ be the true code that is transmitted. Suppose the following conditions are true.
    \begin{enumerate}
        \item $\rank(AX_i)=k_i$.
        \item The rows of the error matrix $E$ have weight at most $\lfloor{\frac{(\delta-1)}{2}\rfloor}$. 
        % \item If $\rank(U)=N$, then  for each $j\in [M]\setminus i$, the matrix $U$ has at least one row which is a codeword in ${\cal C}_i$ but not in ${\cal C}_j$.
    \end{enumerate}
    Then, the $\MDens$ decoder is correct, i.e., it returns the right estimate $\hat{\cal C}_{SD}={\cal C}_i$.
\end{theorem}

\begin{IEEEproof}
When $\rank(AX_i)=k_i$, this implies $\s{AX_i}={\cal C}_i$. For the given $Y$ and the true transmitted  code ${\cal C}_i$, by equation \eqref{eq:decodedrows}, we have $D_i(Y)= AX_i$, as all rows in $E$ have low-weight. Thus, we have 
 \vspace{-0.1cm}   
    {\small \begin{align*}
    &\Delta_{\Dens}(Y,{\cal C}_i)
    =d_s(\s{D_i(Y)},{\cal C}_i)=d_s(\s{AX_i},{\cal C}_i)=d_s({\cal C}_i,{\cal C}_i)=0.
    \end{align*} }

    Now, consider the code ${\cal C}_j$, for some $j\neq i$. We have, by definition,  
    \begin{align*}
    \Delta_{\Dens}&(Y,{\cal C}_j)=d_s(\s{D_j(Y)},{\cal C}_j)\\
    &=\dim(\s{D_j(Y)})+\dim({\cal C}_j)-2\dim(\s{D_j(Y)}\cap {\cal C}_j).
    \end{align*} 
Thus, we observe that if one of the following two conditions are true, then we will have $d_s(\s{D_j(Y)},{\cal C}_j)\geq 1$.
    \begin{enumerate}
        \item $\dim(\s{D_j(Y)})>\dim(\s{D_j(Y)}\cap     {\cal C}_j)$, (or)
        \item $\dim({\cal C}_j)>\dim(\s{D_j(Y)}\cap {\cal C}_j)$. 
    \end{enumerate}
    Thus, if either of the above two conditions are true, the $\MDens$ decoder will be correct. We now show that, indeed, under the conditions in the theorem statement, it will be true that $\dim(\s{D_j(Y)})>\dim(\s{D_j(Y)}\cap {\cal C}_j)$. 

    To see this, note that, as $\s{AX_i}={\cal C}_i$, we know that $AX_i$ contains at least one row (say, $(AX_i)[r_1,:]$) which is a codeword in ${\cal C}_i$ but not in ${\cal C}_j$ (since ${\cal C}_i\subsetneq {\cal C}_j$).   Further, by the condition the rows of $E$ have low weight,  we are assured that the row $D_j(Y)[r_1,:]$ is not a codeword in ${\cal C}_j$ either. To see this, assume the contrary, i.e. that $D_j(Y)[r_1,:]\in{\cal C}_j$. Then, it must be the case that $d_H(D_j(Y)[r_1,:],Y[r_1,:])<\delta/2$, by the definition of $D_j(Y)$ (see \eqref{eq:decodedrows}). Now, by $2)$ we have that $d_H(AX_i[r_1,:],Y[r_1,:])<\delta/2$. Then, by triangle inequality, we see that $d_H(AX_i[r_1,:],D_j(Y)[r_1,:])<\delta$. Since we know that $AX_i[r_1,:]\in{\cal C}_i\setminus {\cal C}_j$, this is a contradiction, by definition of $\delta$ in \eqref{eqn:definitiondelta}. Thus, $D_j(Y)[r_1,:]$ is not a codeword in ${\cal C}_j$. This means $\dim(\s{D_j(Y)})>\dim(\s{D_j(Y)}\cap {\cal C}_j)$. Thus, the $\MDens$ decoder will be correct and this completes the proof. 
    \end{IEEEproof}

%%=============================
\subsection{An Improved Decoder for Large $N$} 
\label{subsec:improveddecoder}

%%=====================================================
%=================================
%
\begin{figure*}[tp]

\begin{center}

\begin{tikzpicture}

% Left image
\node (A) {\includegraphics[width=0.48\linewidth]{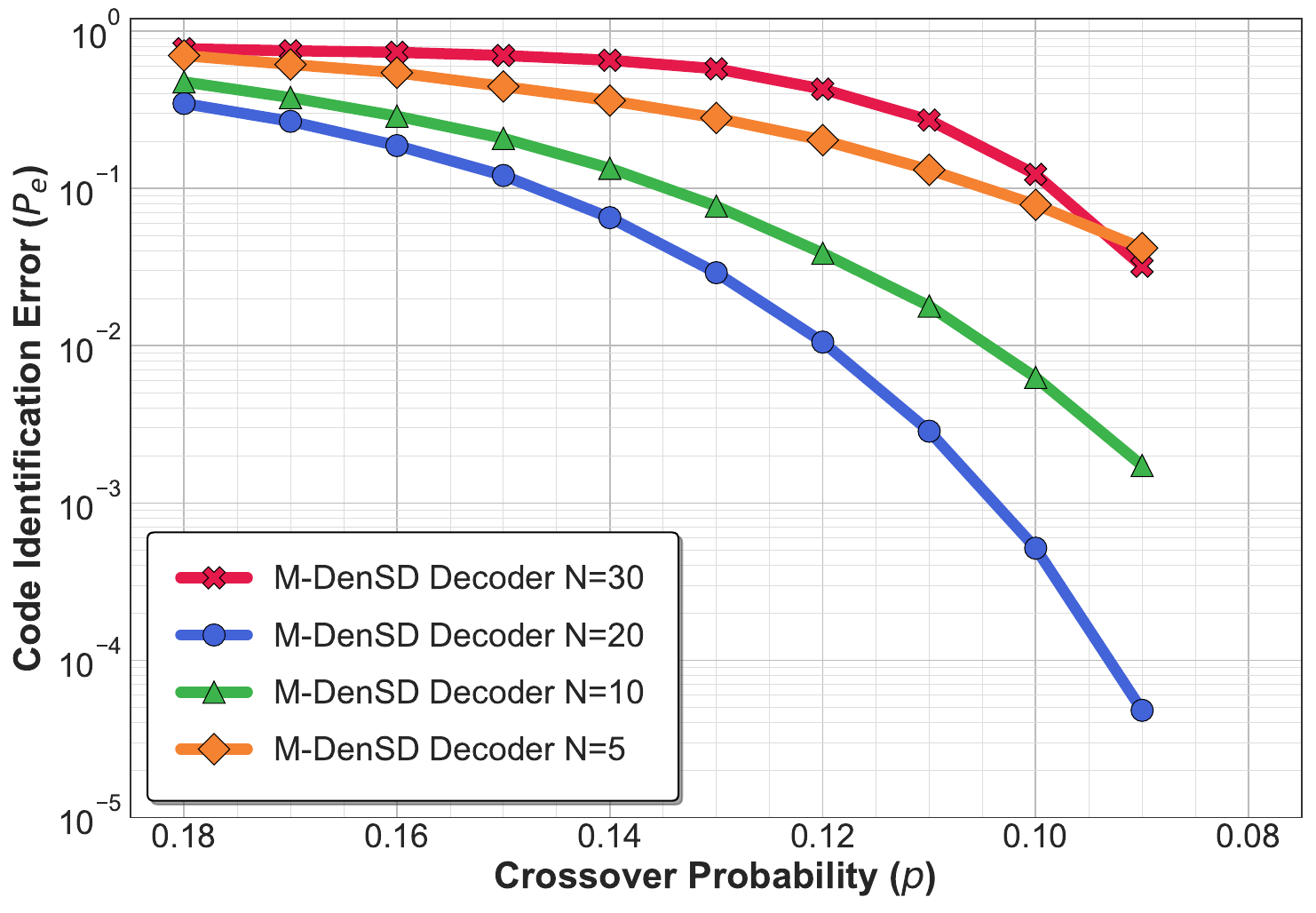}};
\node[below=2mm of A] {\small (a) Performance for different values of $N$};

% Right image
\node[right=6mm of A] (B) {\includegraphics[width=0.48\linewidth]{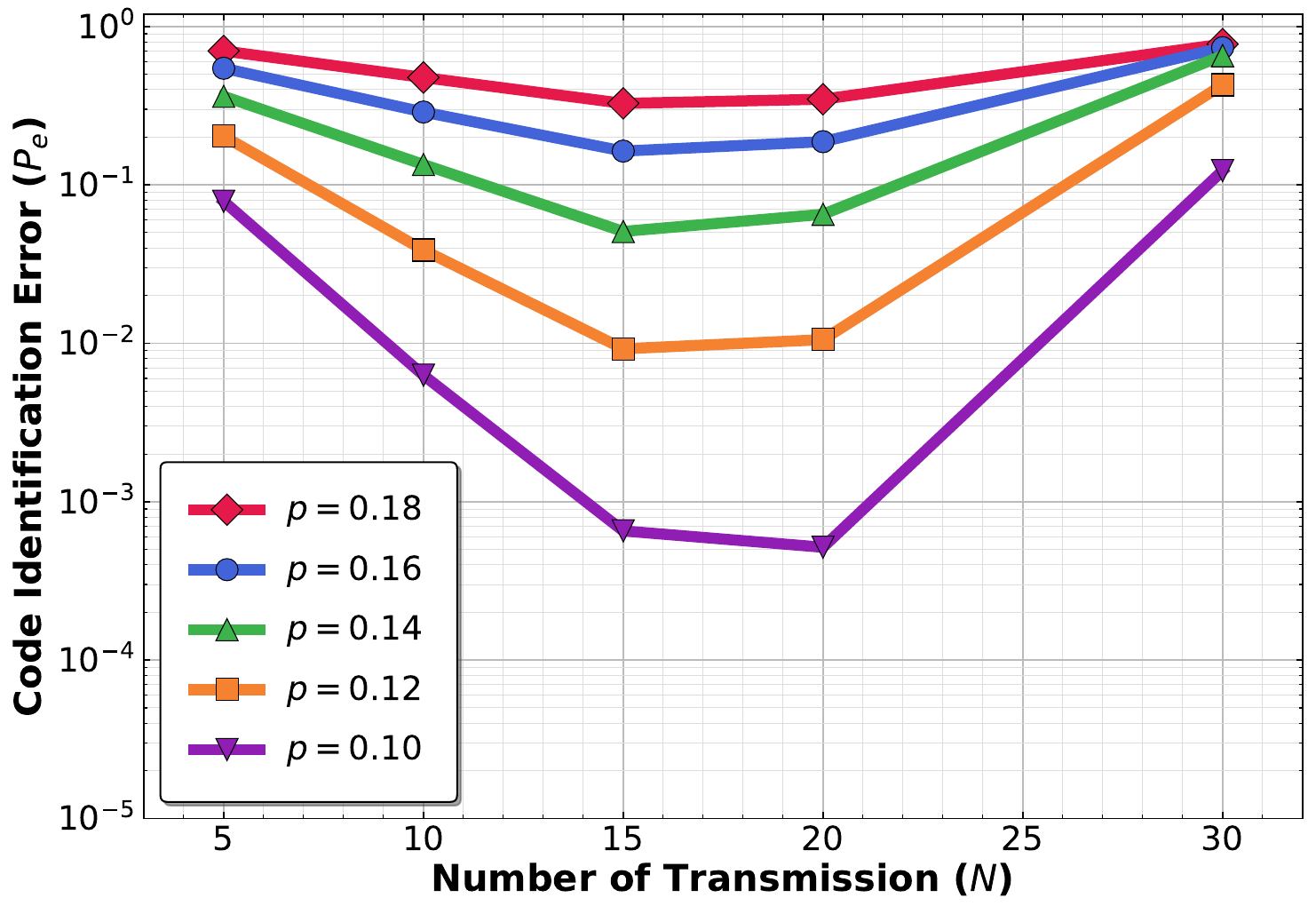}};
\node[below=2mm of B] {\small(b) Performance for different values of $p$};

\end{tikzpicture}

\end{center}

\caption{\small Performance of $\MDens$ decoder indicating the worsening of performance for $N$ beyond some threshold $N^*$ (in this case, $N^*\approx 20$)}

\label{Figure_motivate_Improved_Algo}

\end{figure*}

It is evident that as $N$ increases, the probability that all rows of $E$ have weight at most $\lfloor (\delta-1)/2 \rfloor$ decreases. Due to the accumulation of these higher-weight rows in $E$, the subspace distance $d_s(D_i(Y),{\cal C}_i)$ (see \cref{def:disc}) can increase even for the correct code ${\cal C}_i$, thus leading to identification error. 
As a result, the performance of the $\MDens$ decoder (see \cref{eq: mdd}) plausibly monotonically improves until $N\leq N^*$ for some appropriate $N^*$, and for $N>N^*$ it can potentially worsen.
A numerical evidence of this is illustrated in Fig.~2 for a code family with two codes of dimension $k=10$ and length $30$. It can be seen from Fig.~2(b) that, for this example, $N^*\approx 20$.
However, it appears difficult to characterize this threshold $N^*$, as this involves the interplay of the event of seeing high-weight error vectors which also lead to large subspace distance of $D_i(Y)$ from the true code ${\cal C}_i$.  %However, in %Subsection \ref{subsec:simulations} where we show simulation results, we show some evidence to this phenomenon (see Fig.~\ref{fig:error_vs_N}, 

% Moreover, for larger values of $N$, decoding according to the $\MDens$ Decoder does not guarantee correct identification of the transmitted code.

To overcome this issue as $N$ increases beyond $N^*$, we propose a revised decoding rule, as follows. 
%%%%%
\begin{equation}
\label{eq:subset Decoder}
\hat{\mathcal{C}}_{ISD} =  \arg\min_{\mathcal{C} \in \mathfrak{C}} \left (  \min_{\substack{R \subseteq [N] \\ |R| = N^*}} d_s\big(\s{D(Y)[R,:]}, \mathcal{C}\big)\right).    
\end{equation}
Here $D(Y)[R,:]$ denotes the submatrix of $D(Y)$ consisting of only the rows given by $R$.  We henceforth refer to the decoder in \eqref{eq:subset Decoder} as the \textit{Improved} $\MDens$ decoder. Note that, via this decoder, Theorems \ref{Theorem_MdenSD_Correct} and \ref{Theorem_MSDD_Correct_2} can be easily modified, where the theorems now hold as long as the stated respective conditions in these hold for at least one subset of $N^*$ rows of $AX_i$ and $E$, instead of all rows.

Clearly, the  decoder in \eqref{eq:subset Decoder} is computationally more intensive than the $\MDens$ decoder, as it goes through all possible $R\subset [N]$ such that $|R|=N^*$. However, it can perform better than the $\MDens$ decoder, as it seeks a subset of rows which minimizes the subspace distance to the candidate code. Results in Subsection \ref{subsec:simulations} show that this is indeed  the case. In \cref{eq:subset Decoder}, note that one needs to consider all possible $\binom{N}{N^*}$ submatrices. Since this number could become large, for computational tractability, we consider a small collection of $l\leq \binom{N}{N^*}$ submatrices for the minimization in \cref{eq:subset Decoder} when implementing our decoder. \cref{alg:decoder} summarizes the steps of our implementation.

%In this algorithm, given $Y$ we first find $D(Y)$ (see \cref{eq:decodedrows}) for every $\mathcal{C} \in \mathfrak{C}$. We then consider a matrix formed by  

%\textit{Algorithm for $\MDens$ Decoder}
%
%We now present a general decoding algorithm for code-identification. The decoder takes as input the channel output matrix $Y$ with $N$ rows, and it is assumed that the code family $\mathfrak{C}$ is known to the decoder. For each code $\mathcal{C} \in \mathfrak{C}$, the decoder performs row-wise bounded-distance decoding on $Y$ with respect to $\mathcal{C}$ up to radius $\lfloor (\delta-1)/2 \rfloor$. Specifically, each row of $Y$ is replaced by the nearest codeword of $\mathcal{C}$ if such a codeword exists within the Hamming distance $\lfloor (\delta -1)/2 \rfloor$ otherwise the row is left unchanged. The resulting matrix is denoted by $D(Y)$ and is referred to as the denoised output corresponding to $\mathcal{C}$. Finally, the decoder computes the subspace distance between $\langle D(Y)\rangle$ and $\mathcal{C}$, and outputs the code which minimizes this distance among all the codes in code family $\mathfrak{C}$.
%\tcr{TO-DO: Need to explain why algorithm is considering $N^*$ sized $l\leq \binom{N}{N^*}$ subset.} \tcb{Arti: Done}

\begin{algorithm}[!t]
\caption{Improved $\MDens$ Decoder}
\label{alg:decoder}
\begin{algorithmic}[1]

\State \textbf{Input:} Channel output $Y \in \mathbb{F}_2^{N \times n}$, code family $\mathfrak{C}$, parameters $N^{*}\in [N]$ and integer $l$ such that $\binom{N}{N^*}\geq l\geq 1.$
\State \textbf{Output:} Estimate of transmitted code $\hat{\mathcal{C}}$. 

\State Initialize: $d_{s,\min} \gets \infty$

\ForAll{$\mathcal{C} \in \mathfrak{C}$}

    \State $D(Y) \gets Y$

    \ForAll{$j \in [N]$}
        % \State \tcb{Arti: $\mathbf{c} \gets$ Perform bounded- }     
        \State Attempt bounded-distance decoding (BDD) of $\row{D(Y)}{j}$ into ${\cal C}$, upto distance $\lfloor\frac{\delta-1}{2}\rfloor$. If BDD succeeds, update $\row{D(Y)}{j}$ as the codeword obtained. Else, retain $\row{D(Y)}{j}$ as is. 
        % \If{$d_H(\row{D(Y)}{j}, \mathbf{c}) < \frac{\delta}{2}$}
        %     \State $\row{D(Y)}{j} \gets \mathbf{c}$
        % \Else
        %     \State  $\row{D(Y)}{j} \gets \row{D(Y)}{j}$
        % \EndIf
    \EndFor

    \If{$N \le N^{*}$}
        \State $d_{s,{\cal C}} \gets d_s(\langle D(Y) \rangle, \mathcal{C})$
    \Else
        \State Sample $l$ distinct $N^*$-sized subsets $\{R_1, \dots, R_l\}$ of $[N]$ uniformly at random.
        
        \State $d_{s,{\cal C}} \gets \min_{1 \le i \le l} d_s(\langle {\row{D(Y)}{R_i} \rangle}, \mathcal{C})$
    \EndIf

    \If{$d_{s,{\cal C}} < d_{s,\min}$}
        \State $d_{s,\min} \gets d_{s,{\cal C}}$
        \State $\hat{\mathcal{C}} \gets \mathcal{C}$
    \EndIf

\EndFor

\State \Return $\hat{\mathcal{C}}$

\end{algorithmic}
\end{algorithm}

%----------------------------------
%\subsection{Complexity of $\MDens$ }
\subsubsection*{Complexity of Improved $\MDens$}:
We now analyze the computational complexity of the proposed algorithm. 
%Let $\mathfrak{C}$ be a collection of $M$ linear $(n,k)$ codes.
For each code $\mathcal{C} \in \mathfrak{C}$, the decoder performs two principal operations:
(i) bounded-distance syndrome decoding of selected rows, and
(ii) computation of subspace distances, which reduces to rank computations over the finite field.

The bounded distance decoding step is implemented using  syndrome decoding. Syndrome decoding for a general linear code is NP-hard in the worst case. In this work, we restrict decoding to radius $\lfloor (\delta-1)/2 \rfloor$ and employ a precomputed syndrome lookup table, under which decoding a single length-$n$ vector requires $\mathcal{O}(n(n-k))$ operations, assuming that the look-up is done in constant time. Since the algorithm performs syndrome decoding on $N$ rows,  the total decoding cost per code $\mathcal{C}$ is $\mathcal{O}(N n(n-k))$.

Next, computing the subspace distance of $D(Y)$ from a code ${\cal C}$ involves rank computations of matrices of dimensions at most $(N^*+k)\times n$ (by \eqref{eq:subdist_matrix}). A rank computation of a  matrix of size $a \times b$ requires $\mathcal{O}(ab\min\{a,b\})$ operations. Assuming $N^*+k \ge n$, this is $\mathcal{O}((N^*+k) n^2)$ per subset. Thus, the total rank-computation cost per code is $\mathcal{O}\!\left(l(N^*+k)n^{2}\right)$. Combining both, the total complexity of the proposed decoder is
$
\mathcal{O}\!\left(
M\big(N n(n-k) + l(N^*+k)n^{2}\big)
\right).
$ Expressing $k = Rn$, where $R$ is the code rate, this becomes $\mathcal{O}\!\left( n^{2}\big(MN(1-R) + lN^*\big) + M l R n^{3} \right),$ and for fixed $N^*$ and $l$, the overall complexity scales as $\mathcal{O}\!\left(M l R n^{3}\right).$

\subsection{Bound on the Error Probability of Improved $\MDens$ Decoder}
\label{subsec:errorprobbound}
We now provide a bound on the probability of error of the Improved $\MDens$ decoder, when using an equi-dimension code family $\mathfrak{C}$, where all codes in $\mathfrak{C}$ have dimension $k$. Let $N^*\leq N$ be some value chosen for the definition \eqref{eq:subset Decoder} of the Improved $\MDens$ decoder. Let $\cal R$ denote the set of all $N^*$-sized subsets of $[N]$. Recalling our input-output relationship $Y=AX+E$, consider the following two events: 
\begin{enumerate}
    \item Event ${\cal E}_1$: The set ${\cal R}'\triangleq \{R\in{\cal R}: \w{(\row{E}{i})}< \lfloor (\delta -1)/2 \rfloor, \forall i\in R\}$ is non-empty, and for every such $R\in {\cal R}'$ we have $\rank(\row{A}{R}) < k$. 
    \item Event ${\cal E}_2$: For each $R\in{\cal R}$, we have $\w{(\row{E}{i})} > \lfloor (\delta -1)/2 \rfloor$, for some $i \in R$.
\end{enumerate}
By Theorem \ref{Theorem_MSDD_Correct_2}, the improved $\MDens$ decoder will be correct if the complement of the event ${\cal E}_1\cup {\cal E}_2$ occurs
% \pk{replace old text: `\textit{if the complements of both events ${\cal E}_1$ and ${\cal E}_2$ occur for at least one subset $R$}', with new text `\textbf{if the complement of the event ${\cal E}_1\cup {\cal E}_2$ occurs}'}.  
Thus, the probability of error $\mathsf{P}_e$ of the decoder is bounded as follows. 
\begin{align*}
\mathsf{P}_e 
&\leq P({\cal E}_1\cup {\cal E}_2)\leq P({\cal E}_1)+P({\cal E}_2) 
% \\
% &\overset{(a)}{\le} \mathsf{P}\!\left(\rank(\row{A}{R}) \le k-1\right) \\
% &\quad + \mathsf{P}\!\Big(
% \nexists\ R \subseteq [N]:
%  \forall i \in R\, \w{\row{E}{i}} \le \left\lfloor \frac{(\delta -1)}{2} \right\rfloor
% \Big).
\end{align*}
% where (a) follows from the union bound. Decoding fails if either the selected submatrix of $A$ does not have full rank or submatrix of $E$ has fewer than $N^*$ rows satisfy the bounded-weight condition.
Note that $P({\cal E}_1)\leq \sum_{r=0}^{k-1} \mathsf{P}(\rank(\row{A}{R})=r),$ for any $R\in{\cal R}'$
% \pk{slightly changed: $P({\cal E}_1)\leq \sum_{r=0}^{k-1} \mathsf{P}(\rank(\row{A}{R})=r),$ for any $R\in{\cal R}'$}. 
Now, for a uniformly random binary matrix $A_1 \in \mathbb{F}_2^{N^* \times k}$, the rank distribution satisfies 
$\mathsf{P}(\rank(A_1)=r) \le 2^{-(N^*-r)(k-r)}$, for $0\leq r\leq \min(N,k)$ (see \cite{rankweight}, for arguments based on which we can arrive at this bound). Thus, $P({\cal E}_1)\leq \sum_{r=0}^{k-1}2^{-(N^*-r)(k-r)}.$

Further, denote by $\alpha(p)$ the probability that a row of $E$ has weight more than $\left\lfloor (\delta -1)/2 \right\rfloor$, i.e.,
$$
\alpha(p) = \sum_{i= \left\lfloor \frac{(\delta -1)}{2} \right\rfloor + 1}^{n} 
\binom{n}{i} p^i(1-p)^{n-i}. 
$$
Then, event $\mathcal{E}_2$ occurs if there are fewer than $N^*$ rows with weight $\le \lfloor (\delta-1)/2 \rfloor$. The probability of this is 
$$
P({\cal E}_2)=\sum_{i=N+1-N^{*}}^{N}
\binom{N}{i} \alpha(p)^{i}(1-\alpha(p))^{N-i}. 
$$

Combining the arguments above, we get the upper bound on $\mathsf{P}_e$
%%%%%
\begin{align}
\nonumber \mathsf{P}_e 
\label{eqn:proboferrtighterbound}&\le \sum_{r=0}^{k-1}2^{-(N^*-r)(k-r)} \\
&\quad + \sum_{i=N+1-N^{*}}^{N}
\binom{N}{i} \alpha(p)^{i}(1-\alpha(p))^{N-i}\\
\nonumber
\label{eqn:proboferrorlooserbound}
&\overset{}{\le} k 2^{-(N^* + 1 - k)} \\
&\quad + \sum_{i=N+1-N^{*}}^{N}
\binom{N}{i} \alpha(p)^{i}(1-\alpha(p))^{N-i}
\end{align}
\begin{remark}
    Observe that the first terms in the R.H.S. of both \eqref{eqn:proboferrtighterbound} and \eqref{eqn:proboferrorlooserbound} does not depend on $N$, and is decided by the value of $N^*$ (given $k$). In particular, this first term in each of \eqref{eqn:proboferrtighterbound} and \eqref{eqn:proboferrorlooserbound} \textit{decreases} as $N^*$ increases, for any fixed $N$. 
% \pk{NOTE CORRECTIONS HERE. Previous version was: `Observe that the first term in the R.H.S. of both \eqref{eqn:proboferrtighterbound} and \eqref{eqn:proboferrorlooserbound} does not depend on $N$, and is \textit{minimized} (attaining a constant floor) as $N^*$ increases, for any fixed $N$.'} 
    On the other hand, the second term is the tail of the binomial distribution with parameters $N,\alpha(p)$, and this \textit{increases} 
    as $N^*$ increases, for any fixed $N$. This indicates a `sweet-spot' for $N^*$, for fixed $N$, which was observed via simulations in Subsection \ref{subsec:improveddecoder}. Characterizing this optimizing $N^*$ is an ongoing work. As the mean of the binomial distribution is $N\alpha(p)$, we also observe that choosing $N^*=cN$ for some constant $c<1-\alpha(p)$, the bound on $\mathsf{P}_e$ goes to $0$ as $N\to \infty$. 
    \hfill $\square$    
\end{remark}
%%%%%

%%=====================================================
%%=====================================================
%
\section{Simulation Results}
\label{sec:Simulations}

%\subsection{Algorithm for $\MDens$ Decoder}

\label{subsec:simulations}

We now present the performance of Algorithm \ref{alg:decoder} for randomly generated linear codes.
The simulations conducted in this work consider the simple case of an equi-dimensional code family $\mathfrak{C} = \{\mathcal{C}_1, \mathcal{C}_2\}$ (both with dimension $k$) consisting of two randomly generated linear codes. However, the proposed algorithm is also applicable to codes of different dimensions.
The codes ${\cal C}_1$ and ${\cal C}_2$ are constructed as follows.
Given parameters, $k$ (code dimension), $n$ (blocklength, greater than $k$) and the desired dimension of the intersection $k_{1,2}=\dim({\cal C}_1\cap{\cal C}_2)<k$, we sample
a $(2k-k_{1,2})\times n$ full-row-rank binary matrix uniformly at random. 
The first $k$ rows of this matrix are selected as a basis for $\mathcal{C}_1$,
while the last $k$ rows are selected as a basis for $\mathcal{C}_2$. Next, we compute the (approximate \footnote{It is well known that random linear codes achieve the Gilbert--Varshamov (GV) bound \cite{Gallager1960LDPC} i.e., with high probability, a random linear code of rate $R$ has minimum distance approximately
$n H_2^{-1}(1-R)$, where $H_2^{-1}()$ denotes the inverse binary
entropy function.}) minimum distance
$\delta = d_{\min}(\mathcal{C}_1 + \mathcal{C}_2)$ (see Remark~\ref{remark:delta}). Based on this value, we construct the syndrome decoding table for all error patterns of weight at most $\left\lfloor \delta/2 \right\rfloor$. 

Given the received matrix $Y$ at decoder, we implement a computationally feasible version of the Improved $\MDens$ decoder as provided in Algorithm \ref{alg:decoder}. 

\begin{figure}[ht]
    \centering
    \includegraphics[width= 0.80\linewidth]{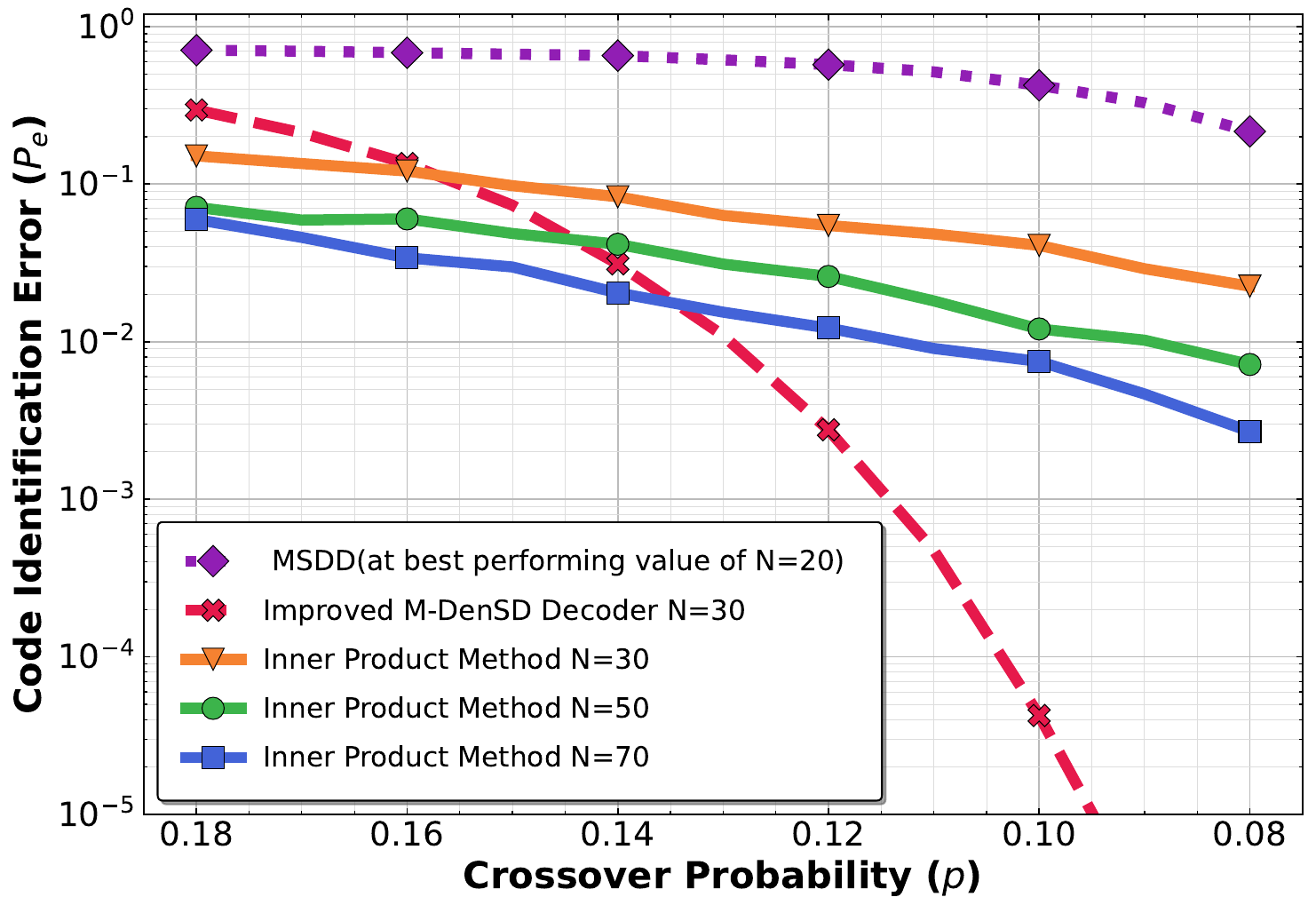}
    \caption{\small Performance comparison: Improved $\MDens$, Inner-Product Method, and MSD decoders ($n= 30, k =10, k_{1,2}=5$)}
    \label{fig:comparison}
\end{figure}

We compare the performance of the Improved $\MDens$ decoder with the well-known  \textit{inner-product method} reported in the literature (See Section \ref{subsetion:lit review}), and also the minimum subspace distance (MSD) decoder (as given in \eqref{eq:MSDD}).  Figure ~\ref{fig:comparison} compares the performance of the three methods for two random linear codes with parameters $n=30,k=10$, and  $k_{1,2} = 5$. Note that, for this case, the minimum distance parameter $\delta\approx \lfloor 0.11n\rfloor\approx 3$, which explains the range of $p$ values in which we see reasonably low error rates. For the inner-product method, we present results for various values of $N$, while for the Improved $\MDens$ algorithm, we only choose $N=30$. The best performance of the MSD decoder is for $N=20$, which is however worse than both of the other methods. As expected, the performance of inner product method improves as $N$ increases. However, the $\MDens$ decoder significantly outperforms the \textit{inner-product method} approach, even at relatively small values of $N =30$, for a considerable range of $p$ values. The performance of the Improved $\MDens$ decoder, for this parameter set, as $N$ increases from $5$ to $30$, is shown in figure \ref{fig:sim_vs_analytical_combined} (b). We see that the slopes of the error rates fall steadily as $N$ grows, indicating increasingly better performance even with a few additional transmitted codewords. 

\usetikzlibrary{positioning}
\begin{figure}[ht]
\centering

\begin{tikzpicture}

% Left image
\node (A) {\includegraphics[width=0.48\linewidth]{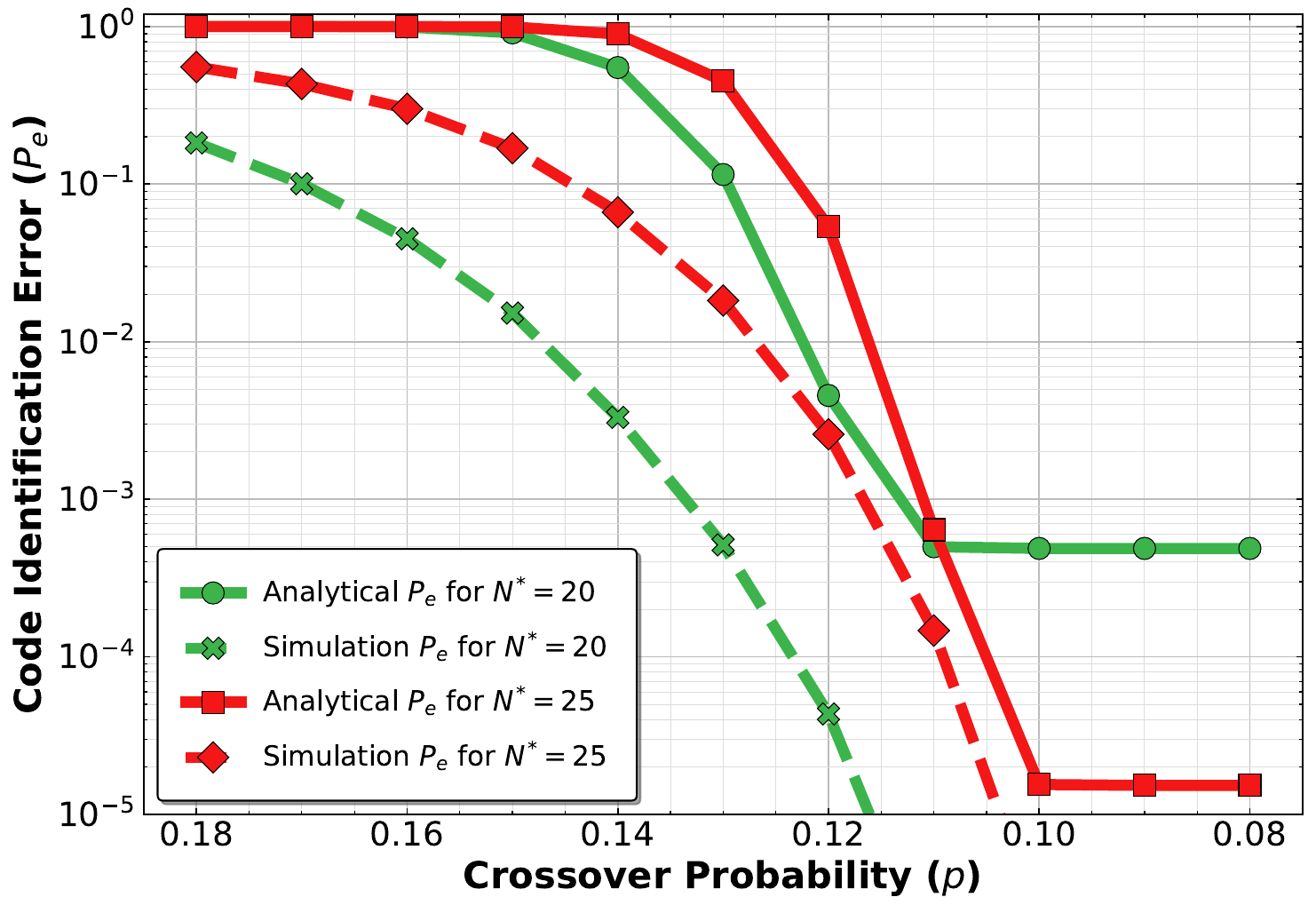}};
\node[below=2mm of A] {\small (a) Simulation vs. Analytical};

% Right image
\node[right=6mm of A] (B) {\includegraphics[width=0.48\linewidth]{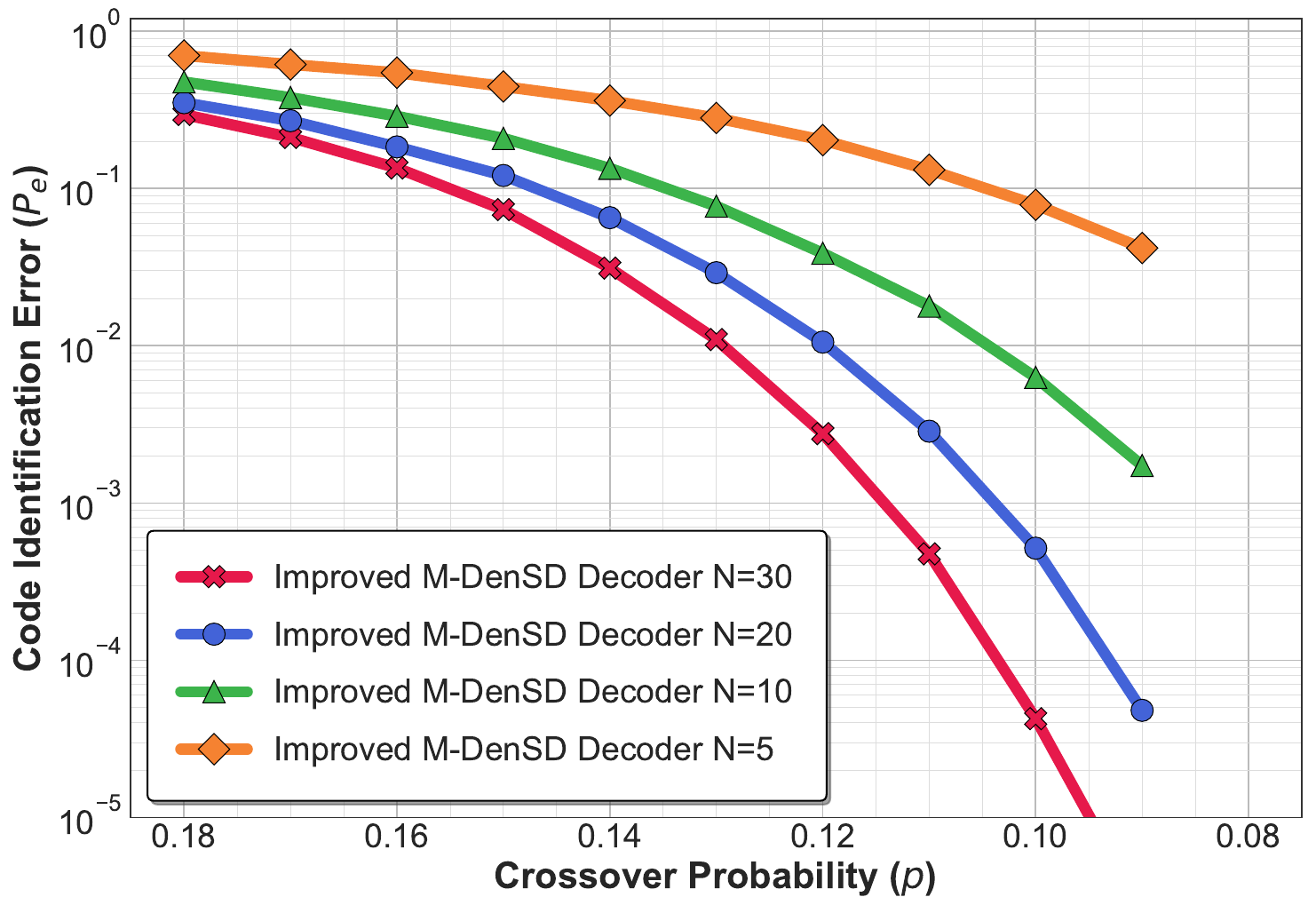}};
\node[below=2mm of B] {\small (b) Performance of Imporved $\MDens$ Decoder};

\end{tikzpicture}

\caption{\small Error probability of Improved $\MDens$ decoder.}
\label{fig:sim_vs_analytical_combined}

\end{figure}

The performance of the Improved $\MDens$ decoder is also shown for codes with lengths around $60$ in Fig. \ref{fig:improvMdensdlong} (a) and Fig. \ref{fig:improvMdensdlong} (b), illustrating the same behaviour. The inner-product method is computationally intensive for these larger length random codes, due to reasons mentioned in Section \ref{subsection_literature_review}. Further, we do not observe the inner-product method or the MSD decoder to have better performance than the Improved $\MDens$ decoder in these cases. Hence, we do not report these findings. 

\usetikzlibrary{positioning}

\begin{figure}[ht]
\centering

\begin{tikzpicture}

% Left figure
\node (A) {\includegraphics[width=0.48\textwidth]{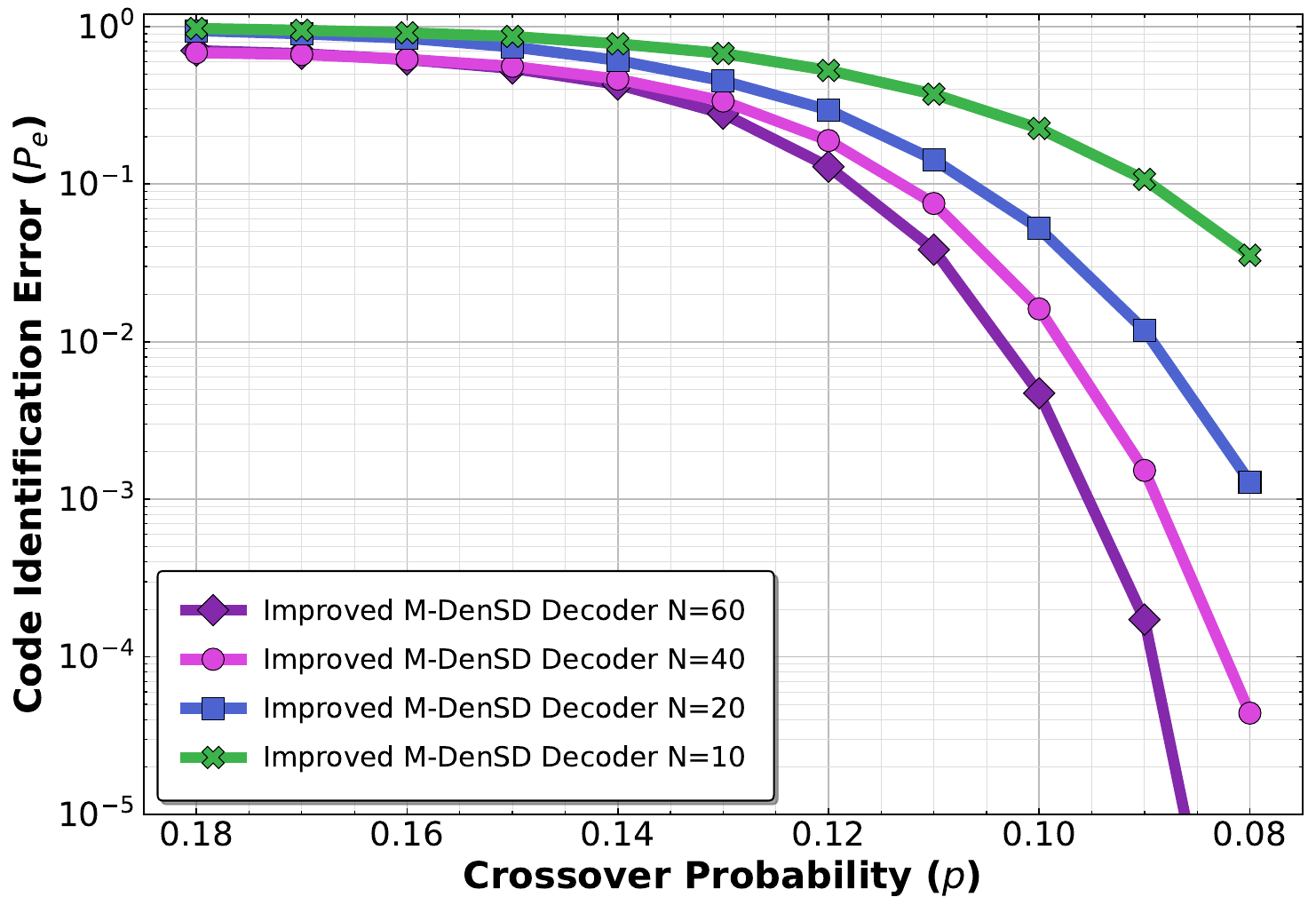}};
\node[below=2mm of A] {\small (a) $n=60,\;k=20,\;k_{1,2}=10$};

% Right figure
\node[right=6mm of A] (B) {\includegraphics[width=0.48\textwidth]{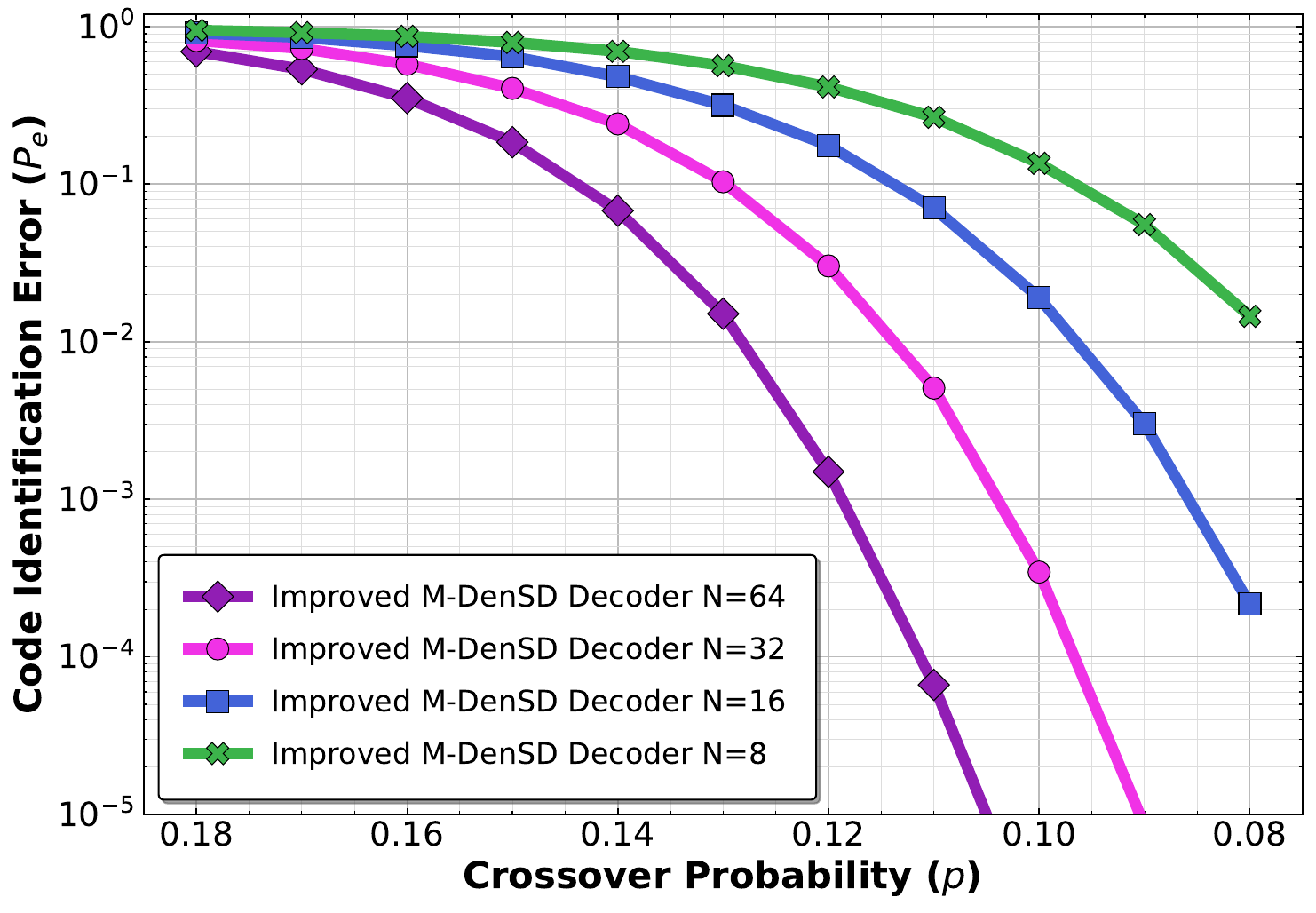}};
\node[below=2mm of B] {\small (b) $n=64,\;k=16,\;k_{1,2}=8$};

\end{tikzpicture}

\caption{\small Performance of Improved $\MDens$ Decoder for different code parameters.}
\label{fig:improvMdensdlong}

\end{figure}

Finally, we remark on Fig. \ref{fig:sim_vs_analytical_combined} (a), which presents a comparison between the error rate of Algorithm \ref{alg:decoder} and the analytical bound on the error-probability of the Improved $\MDens$ decoder, obtained in \eqref{eqn:proboferrtighterbound} in Subsection \ref{subsec:errorprobbound}.  We see that the bound \eqref{eqn:proboferrtighterbound} captures the slope of the error-rate curves closely, until the bound hits a floor due to the first of the two terms in \eqref{eqn:proboferrtighterbound}.

\section{Conclusion and Future Work}
In this work, we studied the problem of blind identification of channel codes under the novel framework of subspace coding and provided a new code-identification method that can be applied to any arbitrary channel codes.
In our approach, to connect the random error of a communication channel with the subspace errors, we introduced an appropriate notion of discrepancy and used it to devise a new code-identification method. We provided analytical guarantees of our method and also verified its performance via simulations.
The performance comparisons we have included in this work are preliminary, covering only the simple methods in the literature and for short length codes. Nevertheless, we believe they are interesting and highlight the effectiveness of subspace-coding based approaches for the blind code-identification problem. 

In the future, it would be interesting to extend the ideas from our subspace coding framework to other related problems such as identifying the code-length, achieving synchronization of the data, and blind code-identification within particular code families. Further, extending our approach for other communication channels such as the binary erasure channel and the AWGN channel is also of interest.

Finally, our new techniques exploit a natural connection between the code-identification problem (over a communication channel) and the subspace coding problem (over an operator channel). 
We would like to highlight here that, in the literature, until now these two research threads have been studied separately. We believe that establishing a connection between them is interesting on its own.

% \pk{Maybe added: The performance comparisons we have included in this work are preliminary, covering only the simple methods in the literature and for short length codes. Nevertheless, we believe they are interesting and highlight the effectiveness of subspace-coding based approaches for the blind code-identification problem. 
% We hope to perform comparison for longer codes in future work, by utilizing optimized versions of the algorithms for finding low-weight dual codewords.
% }

%%%%%%
%% To balance the columns at the last page of the paper use this
%% command:
%%
%\enlargethispage{-1.2cm} 
%%
%% If the balancing should occur in the middle of the references, use
%% the following trigger:
%%
%\IEEEtriggeratref{7}
%%
%% which triggers a \newpage (i.e., new column) just before the given
%% reference number. Note that you need to adapt this if you modify
%% the paper.  The "triggered" command can be changed if desired:
%%
%\IEEEtriggercmd{\enlargethispage{-20cm}}
%%
%%%%%%

%%%%%%
%% References:
%% We recommend the usage of BibTeX:
%%
%\bibliographystyle{IEEEtran}
%\bibliography{definitions,bibliofile}
%%
%% where we here have assumed the existence of the files
%% definitions.bib and bibliofile.bib.
%% BibTeX documentation can be obtained at:
%% http://www.ctan.org/tex-archive/biblio/bibtex/contrib/doc/
%%%%%%

%% Or you use manual references (pay attention to consistency and the
%% formatting style!):
% \newpage 
\bibliographystyle{ieeetr}

\bibliography{ISIT_2026}

\end{document}